# The Orbit and Dynamical Evolution of the Chelyabinsk Object

Vacheslav V.EMEL'YANENKO[1*], Sergey A.NAROENKOV[1], Peter JENNISKENS[2,3], and Olga P.POPOVA[4]

[1] Institute of Astronomy of the Russian Academy of Sciences, 48 Pyatnitskaya, Moscow, 119017, Russia.

[2] SETI Institute, Carl Sagan Center, Mountain View, CA 94043, USA.

[3] NASA Ames Research Center, Moffett Field, CA 94035, USA.

[4] Institute for Dynamics of Geospheres of the Russian Academy of Sciences, 38 Leninsky Prospect, Bldg. 1, Moscow, 119334, Russia.

*Corresponding author. Email: vvemel@inasan.ru

## Abstract

**Abstract** – The orbit of the Chelyabinsk object is calculated, applying the least-squares method directly to astrometric positions. The dynamical evolution of this object in the past is studied by integrating equations of motion for particles with orbits from the confidence region. It is found that the majority of the Chelyabinsk clones reach the near-Sun state. 67 percent of these objects have collisions with the Sun for 15 Myr in our numerical simulations. The distribution of minimum solar distances shows that the most probable time for the encounters of the Chelyabinsk object with the Sun lies in the interval from -0.8 Myr to -2 Myr. This is consistent with the estimate of a cosmic ray exposure age of 1.2 Myr (Popova et al 2013). A parent body of the Chelyabinsk object should experience strong tidal and thermal effects at this time. The possible association of the Chelyabinsk object with 86039 (1999 NC43) and 2008 DJ is discussed.

## Introduction

The Chelyabinsk event on 15 February 2013 was the largest airburst of an asteroid on the Earth atmosphere since the Tunguska phenomenon in 1908. The event occurred unexpectedly over the densely populated Chelyabinsk region. According to official reports, more than 7,320 buildings were affected and 1,613 people asked for medical assistance, most hurt by falling glass. 112 people were hospitalized, 2 in serious condition. This event is the first modern airburst to provide measurable data on how an asteroid impact can cause damage and injuries.

From the astronomical point of view, the Chelyabinsk event gives a large amount of observations for detailed investigations of dynamical and physical properties of the celestial body. The general description of the first results was given in (Emel'yanenko et al., 2013). More comprehensive analysis was done in works (Popova et al., 2013; Borovicka et al., 2013; Brown et al., 2013). The origin of this object was among topics of these papers.

In this work, we present results of our calculations of the orbit and dynamical evolution of the Chelyabinsk object.

## Orbit

Usually methods of the meteor astronomy are applied in calculating the pre-atmospheric meteorioid orbit. At presen,t there are different well-developed approaches in this direction (*e.g.*, Jenniskens et al., 2011; Borovicka et al., 2013). The usual manner in determining the pre-atmospheric orbit is applying Schiaparelli's equation to the calculated radiant and velocity at infinity, and this introduces additional simplifications in the description of the motion. Such an approach is good enough in the case of meteors, but there is no need to use these simplifications in the case of Chelyabinsk with extensive observational data.

Instead, the pre-atmospheric orbit was calculated directly from a least-squares fit to the astrometric data using methods common in calculations of asteroid orbits. Each direction to the fireball front edge from one of the video stations is regarded as one measurement of the meteoroid position from that site. A least-squares fit solution to those directions is then applied to calculate the osculating orbital elements.

This method gives directly not only orbital elements but also their errors. In addition, it allows us to generate possible orbits from the confidence region, using the correlation matrix, to study features of the dynamical evolution. The only difficulty with this method is the need to have an initial orbit that is accurate enough for the convergence of the iteration process of improving the orbit. Therefore, some preliminary orbit is useful here. We take such an initial orbit from the paper (Popova et al., 2013).

We take account of gravitational attraction of the Sun, all planets, and the Moon. In addition, we add the negative term due to the atmospheric drag in the form $-A\rho_A V\vec{V}$ to the equation of acceleration, where $\rho_A$ is the atmosphere density, $\vec{V}$ is the velocity of the object relative to the atmosphere. For the entire observed meteor phenomenon, the coefficient $A$ may be a very complicated function of time due to action of ablation and fragmentation associated with specific properties of the object. Therefore, in our orbit determination 1) we try to use observations in the upper atmosphere where the interaction with the atmosphere and consequently the changes of $A$ are smaller than at the stage of the main disruption (Popova et al., 2013); 2) we find some average constant coefficient $\tilde{A}$ that gives the best fit of observations.

The observational data for the Chelyabinsk object were taken through the internet. They include videos with direct images and shadows from the moving fireball. These videos were calibrated during our two-week field investigation (March 9 – 24, 2013) and additional requests to locals Gennadij Ionov and Artem Burdanov, by taking star background images and Sun-generated shadows at know times. The detailed discussion of observations for the Chelyabinsk object was done in the paper (Popova et al., 2013). For the present paper, we take astrometric positions (right ascension α and declination δ) estimated from the five best videos made in Korkino, Pervomayskiy, Beloretsk, Snezhinsk and Kamensk-Uralskiy and shadow measurements in Chebarkul. These observations were checked carefully by P.Jenniskens on systematic errors in (Popova et al., 2013). Our numerical experiments show that the best fit of observations with $t$<03:20:32.2 UT (a peak in brightness on the lightcurve (Popova et al., 2013)) takes place at $\tilde{A} = 0$. After this moment the absolute value of $\tilde{A}$ increases rapidly. These calculations are consistent with the conclusion made in (Popova et al., 2013) that no significant deceleration was measured above 25 km altitude. Therefore, we determine the orbit of the Chelyabinsk object from the least-squares solution for the observations with $t$<03:20:32.2 UTC at $\tilde{A} = 0$. This orbit (Table 1) fits observations with a standard deviation of 2265″ for the residuals $\Delta\alpha\cos\delta$ and $\Delta\delta$. Note that each orbital element is expressed in more significant digits than justified by the formal uncertainty in the solution, because the relatively accurately observed position of the meteoroid dictates a higher accuracy for any given orbit solution than implied by the uncertainty in entry speed and velocity at that point. Differences in orbital elements between this orbit and the orbits determined in (Popova et al., 2013; Borovicka et al., 2013) are in the range of statistical uncertainties.

Table 1. Pre-atmospheric orbital elements for the Chelyabinsk object (Equinox J2000).

| Orbital elements | |
|---|---|
| $a$, AU | 1.8804986±0.068 |
| $e$ | 0.608988648±0.017 |
| $\omega$(deg) | 108.926276±0.536 |
| $\Omega$ (deg) | 326.445884±0.002 |
| $i$ (deg) | 5.938266±0.427 |
| $M_0$ (deg) | 16.872985±1.069 |
| $T_0$ | 2013.02.15.0 TT |

For comparison, if we use the longer interval of observations up to $t$=03:20:34.5 UTC, then the best fit of observations with a standard deviation of 2458″ takes place at $\tilde{A} = 9.3 \times 10^{-5}$ m²/kg for the following orbital elements: $a$=1.896±0.044, $e$=0.613±0.010, $\omega$=108.905±0.461, $\Omega$=326.446, $i$=5.879±0.277, $M_0$=16.634±0.692. These orbital elements differ only slightly from values in Table 1. The formal errors are smaller than in Table 1. But the real process of deceleration after the explosion is evidently much more complicated than that described in our model with constant $\tilde{A}$, and the meteor image is fuzzier. These circumstances can lead to large systematic errors. Therefore, we give preference to the orbit presented in Table 1 and apply it for the further study. Fig. 1 shows observed and calculated azimuths and elevations for observations used in the orbit determination.

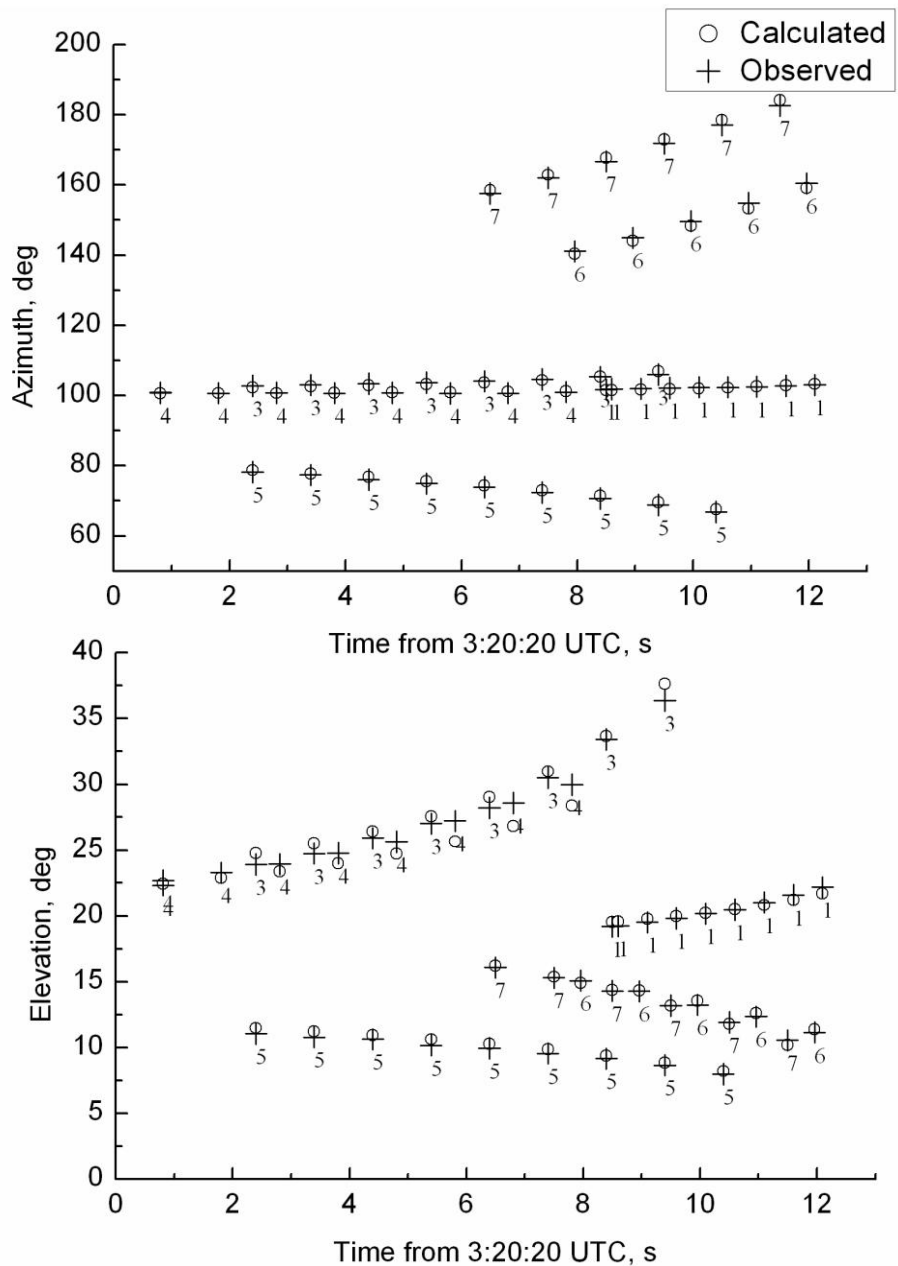

Fig. 1. Observed and calculated angles for observations used in the orbit determination (1 – Chebarkul, 3 – Korkino, 4- Pervomayskiy, 5 – Beloretsk, 6 – Sneginsk, 7 – Kamensk-Uralskiy).

## Long-term evolution

In the case of the Chelyabinsk object, we have a unique combination of numerous video registrations allowing us to calculate a relatively precise orbit with multiple recovered meteorites whose properties can be studied in laboratories. Therefore, to learn more about the origin of Chelyabinsk, it is important to compare dynamical behaviour of the object in the past with data obtained from cosmochemistry analysis.

The general conclusion is that Chelyabinsk originates from the main belt (Popova et al., 2013) following ideas of the paper by Bottke et al. (2002) that almost all near-Earth objects come from this region. From results of this paper, the probability can be estimated that a near-Earth object with a given orbit comes from a certain region of the main belt. Usually the time of the evolution from the main belt to the Chelyabinsk-like orbit is very long (more than ~10 Myr), whereas the age of the Chelyabinsk object as a separate body is much shorter: ~1.2 Myr based on the cosmogenic noble gas analysis (Popova et al., 2013). Thus, it makes sense to look for possible mechanisms of the Chelyabinsk body formation in near-Earth space.

Borovicka et al. (2013) suggested that the Chelyabinsk asteroid originated from the 2.2-km-diameter near-Earth asteroid 86039 (1999 NC43) that has a very similar orbit. However, many questions arise concerning various aspects of this assumption based now only on approximate similarity of orbital elements. First, the investigation was done on the 2000 year interval in the past, and the required ejection velocity is about 1-2 km/s. It is not clear what physical mechanism can eject ~20-m body with such a large velocity. We have no observational evidence of either other bodies or meteoroids created by this recent event. There are no traces of such a recent catastrophic phenomenon among cosmogenic data for the Chelyabinsk meteorite either (Popova et al., 2013).

We studied the long-term evolution of the Chelyabinsk object on the basis of our calculated orbit. Since the orbital elements listed in Table 1 have uncertainties, we investigated how the results of our numerical simulations depend on orbital errors. For this purpose, we considered a large set of initial orbits from the confidence region, using the covariance matrix. These orbits were integrated back for 15 Myr taking account of perturbations from all planets. The dynamical evolution of test particles was calculated using the symplectic integrator (Emel'yanenko, 2007). Particles were removed from integrations when they collide with planets or the perihelion distance $q < 0.005$ AU or $a > 50$ AU.

At the beginning we integrated 430 orbits. Then we doubled the number of initial orbits in order to check that qualitative and statistical results of our integrations for 15 Myr are reliable. For 860 initial orbits, 41 particles formally collide with planets, 53 particles become long-period orbit objects with $a > 50$ AU, 572 particles reach orbits with $q < 0.005$ AU. The other particles move mainly in orbits of near-Earth objects until the end of integrations (we have found only 6 particles with $q > 1.3$ AU after 15 Myr). We do not regard the ejection from planets as a realistic scenario for the origin of Chelyabinsk, and the formal collisions of few particles with planets is caused by orbital uncertainties. From the laboratory analysis of recovered meteorites (Popova et al., 2013), Chelyabisk is an LL5 ordinary chondrite. Ordinary chondrites are linked to S-type asteroids (Binzel et al., 1996; Nakamura et al., 2011); therefore, there is no indication from recovered meteorites that this object originates from long-period comets either. It seems that the most important result of this investigation is the large probability that the Chelyabinsk object was near the Sun in the past.

67 percent of objects have collisions with the Sun for 15 Myr in our numerical integrations. The solar tide, thermal stresses and interaction with the solar atmosphere are expected to be severe for objects passing within the solar radius. The degree of destruction due to these effects is uncertain, it depends on many parameters of a particular body (e.g., Holsapple and Michel, 2008; Shestakova and Tambovtseva, 1997; Brown et al., 2011). The Chelyabinsk parent asteroid can be intensely affected by heating and tidal processes not only at collisions, but also at close approaches to the Sun. Fig. 2 shows the cumulative fraction of particles reaching $q<0.1$ AU in the past (the difference in numbers of collisions and passages with $q<0.1$ AU is only a few percent for 15 Myr).

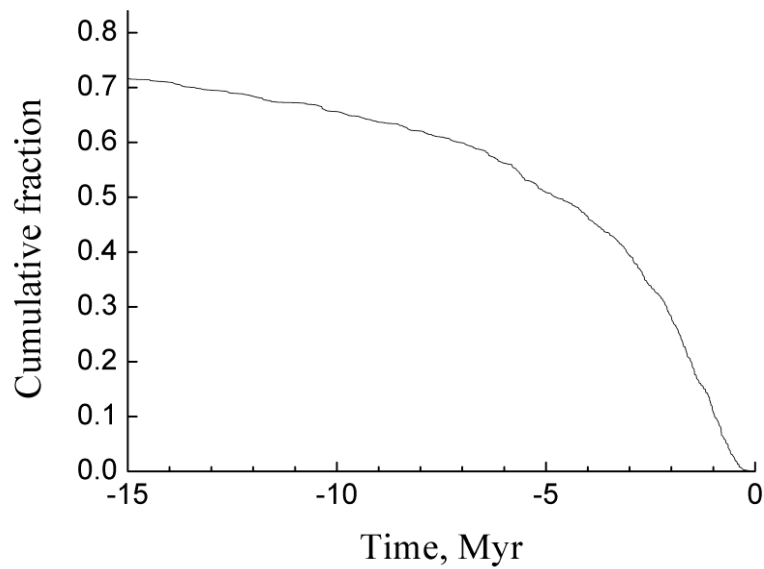

Fig. 2. The cumulative fraction of particles reaching $q<0.1$ AU in the past.

More than 50 percent of particles reach the near-Sun state in 5 Myr. The particles become near-Sun objects most frequently in the time interval from 0.8 Myr to 2 Myr (the cumulative function has the steepest slope in this range). It is remarkable that this is consistent with the estimate of a cosmic ray exposure age of 1.2 Myr (Popova et al., 2013). It is natural to assume that a larger parent body could have been disrupted at Sun-grazing conditions. The recovered fragments of the Chelyabinsk meteorite contain significant portions of shock blackened material and melt veins (Galimov et al., 2013; Kohout et al., 2014; Ozawa et al., 2014) that could be produced in the Sun atmosphere (a temperature of the shock-melt vein matrix formation is estimated over 1700-2000º C in (Ozawa et al., 2014)).

We have not found any case of transition to the main belt orbit. This does not reject the idea that the Chelyabinsk object originated from the main belt asteroid because the likelihood of such findings by backward integrations is extremely small due to chaotic dynamics. The dynamics of near-Earth objects is strongly chaotic, mainly due to occasional close encounters with planets (e.g., Milani et al., 1989; Freistetter, 2009). An example of changes of $a$ and $q$ is shown in Fig. 3 for a selected particle approaching to the Sun near $t$=-1.2 Myr. This particle has many close encounters with terrestrial planets (the most prominent is the encounter with Venus of 0.0006 AU at $t$=-0.73 Myr). Therefore, orbits of individual objects cannot be followed exactly for very long time scales (the mean time it takes for an asteroid to evolve to a Chelyabinsk-like orbit after first entering the near-Earth region is estimated as 3-25 Myr, depending on a source in the main belt (Popova et al., 2013)). Only qualitative and statistical information about the dynamical behaviour can be provided by numerical integrations in this case. In particular, our results show that the near-Sun states are frequent for objects with the Chelyabinsk-type dynamical features on their way to the present near-Earth orbit. This should play an important role in creating a surface and internal structure of the Chelyabinsk body.

The decrease of $q$ for the Chelyabinsk clones is mainly associated with a secular resonance $\nu_6$ with Saturn. Fig. 4 shows the changes of the quantity $\nu_6 = \pi - \pi_s$ for the same particle as in Fig. 3 ($\pi$ and $\pi_s$ are longitudes of perihelion for the particle and Saturn, correspondingly).

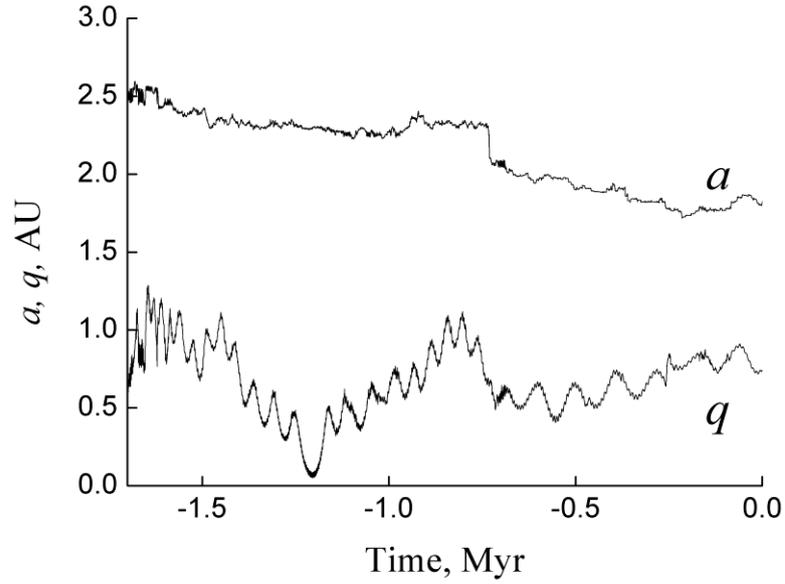

Fig. 3. An example of changes of *a* and *q* for a selected particle approaching to the Sun near *t*=1.2 Myr.

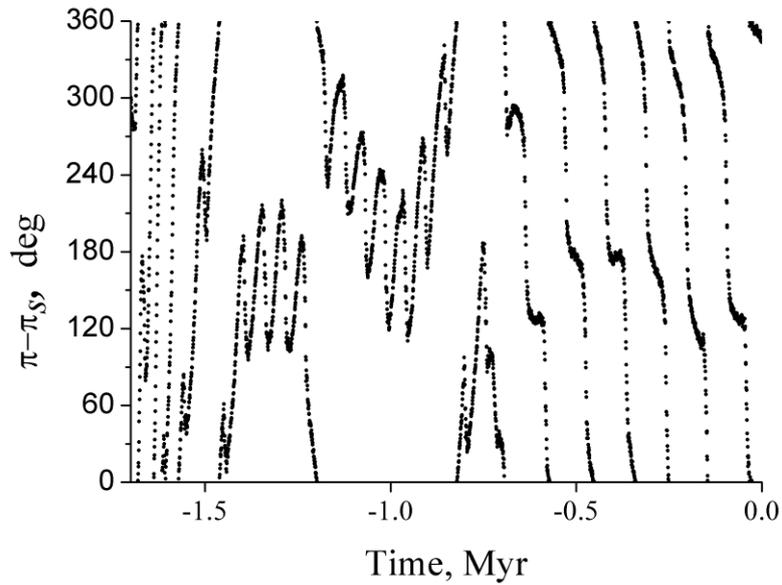

Fig. 4. Changes of $\nu_6 = \pi - \pi_s$ for the same particle as in Fig. 3.

**Comparison with the evolution of 86039 (1999 NC43) and 2008 DJ**

Our investigations have confirmed the conclusion of (Borovicka et al., 2013) that the asteroids 86039 (1999 NC43) and 2008 DJ have currently the lowest dissimilarity criteria *D* (Southworth and Hawkins, 1963) relative to the Chelyabinsk object. We have studied the dynamical evolution of these objects in the same way as for the Chelyabinsk object. Fig. 5 shows that the largest frequency to become a near-Sun object for the 2008 DJ clones takes place between 1.0 and 1.2 Myr. While 2008 DJ has 41% chance to reach $q<0.1$ before 2 Myr in the past, this is equal only 13 % for 86039 (1999 NC43) (28 % for the Chelyabinsk object). Thus from the standpoint of the Chelyabinsk object origin near the Sun, 2008 DJ is preferable to 86039 (1999 NC43) as a parent body.

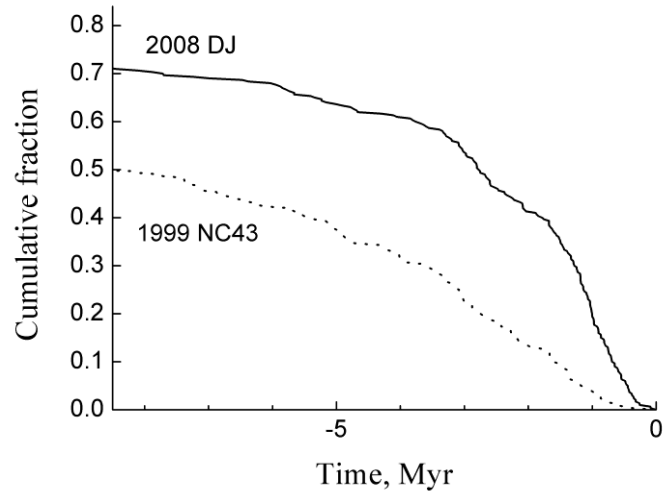

Fig.5. The cumulative fraction of particles reaching $q<0.1$ AU in the past for 1999 NC43 and 2008 DJ.

**Conclusions**

We have determined the orbit of the Chelyabinsk object applying the least-squares method directly to astrometric positions. This has allowed us to study the dynamical evolution of this object in the past, integrating equations of motion for particles with orbits from the confidence region. We have found that the majority of the Chelyabinsk clones reach the near-Sun state. The distribution of minimum solar distances shows that the most probable time for the encounters of the Chelyabinsk object with the Sun lies in the interval from -0.8 Myr to -2 Myr. This is consistent with the estimate of a cosmic ray exposure age of 1.2 Myr (Popova et al., 2013). A parent body of the Chelyabinsk object could be disrupted due to the strong solar tide, thermal stresses and interaction with the solar atmosphere at Sun-grazing conditions. The comparison with the evolution of 86039 (1999 NC43) and 2008 DJ shows that 2008 DJ is preferable to 86039 (1999 NC43) as a parent body.

*Acknowledgments* - We appreciate the referee's careful reading of the paper and helpful comments which led to significant improvements.

# REFERENCES

Binzel R.P., Bus S.J., Burbine T.H., Sunshine T.M. 1996. Spectral properties of 2006 near-Earth asteroids: evidence for sources of ordinary chondrite meteorites. *Science* **273**:946-948.

Borovička J., Spurný P., Brown P., Wiegert P., Kalenda P., Clark D., Shrbený L. 2013. The trajectory, structure and origin of the Chelyabinsk asteroidal impactor. *Nature* **503**:235-237.

Bottke W. F., Morbidelli A., Jedicke R., Petit J.-M., Levison H. F., Michel P., Metcalfe T. S. 2002. Debiased orbital and absolute magnitude distribution of the Near-Earth Objects. *Icarus* **156**:399–433.

Brown J.C., Potts H.E., Porter L.J., Chat G.L. 2011. Mass loss, destruction and detection of Sun-grazing and –impacting cometary nuclei. *Astronomy & Astrophysics* **535**:A71.

Brown P. G., Assink J. D., Astiz L., Blaauw R., Boslough M. B., Borovička J., Brachet N., Brown D., Campbell-Brown M., Ceranna L., Cooke W., de Groot-Hedlin C., Drob D. P., Edwards W., Evers L. G, Garces M., Gill J., Hedlin M., Kingery A., Laske G., Le Pichon A., Mialle P., Moser D. E., Saffer A., Silber E., Smets P., Spalding R. E., Spurný P., Tagliaferri E., Uren D., Weryk R. J., Whitaker R., Krzeminski, Z. 2013. A 500-kiloton airburst over Chelyabinsk and an enhanced hazard from small impactors. *Nature* **503**: 238-241

Emel'yanenko V. V., Popova O. P., Chugai N. N., Shelyakov M. A., Pakhomov Yu. V., Shustov B. M. , Shuvalov V. V., Biryukov E. E., Rybnov Yu. S. , Marov M. Ya. , Rykhlova L. V., Naroenkov S. A., Kartashova A. P. , Kharlamov V. A., Trubetskaya I. A. 2013. Astronomical and Physical Aspects of the Chelyabinsk Event (February 15, 2013). *Solar System Research* **47**:240-254.

Emel'yanenko V.V. 2007. A method of symplectic integrations with adaptive time-steps for individual Hamiltonians in the planetary N-body problem. *Celestial Mechanics and Dynamical Astronomy* **98**:191-202.

Freistetter F. 2009. Fuzzy characterization of near-Earth-asteroids. *Celestial Mechanics and Dynamical Astronomy* **104**:93-102.

Galimov E.M., Kolotov V.P., Nazarov M.A., Kostitsyn Yu.A., Kurbakova I.V., Kononkova N.N., Roshchina I.A., Alexeev V.A., Kashkarov L.L., Badyukov D.D., Sevast'yanov V.S. 2013. Analytical results for the material of the Chelyabinsk meteorite. *Geochemistry International* **51**: 522-539.

Holsapple K.A., Michel P. 2008. Tidal disruptions II. A continuum theory for solid bodies with strength, with applications to the Solar System. *Icarus* **193**:283-301.

Jenniskens P., Gural P. S., Dynneson L., Grigsby B. J., Newman K. E., Borden M., Koop M., Holman D. 2011. CAMS: Cameras for Allsky Meteor Surveillance to establish minor meteor showers. *Icarus* **216**:40–61.

Kohout T., Gritsevich M., Grokhovsky V.I., Yakovlev G.A., Haloda J., Halodova P., Michallik R.M., Penttilä A., Muinonen K. 2014. Mineralogy, reflectance spectra, and physical properties of the Chelyabinsk LL5 chondrite - Insight into shock-induced changes in asteroid regoliths. *Icarus* 228:78-85.

Milani A., Caprino M., Hahn G., Nobili A.M. 1989. Dynamics of planet-crossing asteroids: classes of the orbital behavior. *Icarus.* **78**:212-269.

Nakamura T., Noguchi T., Tanaka M., Zolensky M.E., Kimura M., Tsuchiyama A., Nakato A., Ogami T., Ishida H., Uesugi M., Yada T., Shirai K., Fujimura A., Okazaki R., Sandford S.A., Ishibashi Y., Abe M., Okada T., Ueno M., Mukai T., Yoshikawa M., Kawaguchi J.

2011. Itokawa dust particles: a direct link between S-type asteroids and ordinary chondrites. *Science* **333**:1113-1116.

Ozawa S., Miyahara M., Ohtani E., Koroleva O.N., Ito Y., Litasov K.D., Pokhilenko N.P. 2014. Jadeite in Chelyabinsk meteorite and the nature of an impact event on its parent body. Nature Scientific Reports 4:id. 5033.

Popova O. P., Jenniskens P., Emel'yanenko V., Kartashova A., Biryukov E., Khaibrakhmanov S., Shuvalov V., Rybnov Y., Dudorov A., Grokhovsky V. I., Badyukov D. D., Yin Q.-Z., Gural P. S., Albers J., Granvik M., Evers L. G., Kuiper J., Kharlamov V., Solovyov A., Rusakov Y. S., Korotkiy S., Serdyuk I., Korochantsev A. V., Larionov M. Y., Glazachev D., Mayer A. E., Gisler G., Gladkovsky S. V., Wimpenny J., Sanborn M. E., Yamakawa A., Verosub K., Rowland D. J., Roeske S., Botto N. W., Friedrich J. M., Zolensky M., Le L., Ross D., Ziegler K., Nakamura T., Ahn I., Lee J. I., Zhou Q., Li X.-H., Li Q.-L., Liu Y., Tang G.-Q., Hiroi T., Sears D., Weinstein I. A., Vokhmintsev A. S., Ishchenko A. V., Schmitt-Kopplin P., Hertkorn N., Nagao K., Haba M. K., Komatsu M., and Mikouchi T. 2013. Chelyabinsk airburst, damage assessment, meteorite recovery, and characterization. *Science* **342**:1069–1073.

Shestakova L. I., Tambovtseva L.V. 1997. The thermal destruction of solids near the Sun. *Earth, Moon, and Planets* **76**:19-45.

Southworth R. B., Hawkins G. S. 1963. Statistics of meteor streams. *Smithsonian Contributions to Astrophysics* **7**:261–285.